\documentstyle[12pt,epsfig]{article}
\begin{document}

\begin{flushright}
{\bf hep-ph/0103242}
\end{flushright}

\vspace{0.4cm}

\begin{center}
{\Large\bf How to Describe Neutrino Mixing and CP Violation}
\end{center}

\vspace{0.2cm}
\begin{center}
{\bf Harald Fritzsch} \\
{\small\sl Sektion Physik, Universit$\sl\ddot{a}$t M$\sl\ddot{u}$nchen,
D--80333 M$\sl\ddot{u}$nchen, Germany}
\end{center}

\begin{center}
{\bf Zhi-zhong Xing
\footnote{Electronic address: xingzz@mail.ihep.ac.cn}} \\
{\small\sl Institute of High Energy Physics, P.O. Box 918 (4),
Beijing 100039, China}
\end{center}

\vspace{2.5cm}

\begin{abstract}
We present a classification of possible parametrizations of the
$3\times 3$ lepton flavor mixing matrix in terms of the rotation and
phase angles. A particular parametrization, which is most convenient
to describe the observables of neutrino oscillations and that of the
neutrinoless double beta decay, is studied in detail. The possibility 
to determine the Dirac- and Majorana-type CP-violating
phases are also discussed.
\end{abstract}

\newpage

The accumulating evidence for atmospheric and solar neutrino oscillations
strongly suggests that neutrinos are massive and lepton flavors are 
mixed \cite{SK}.
In general the flavor mixing among $N$ different lepton families is 
described by a $N\times N$ unitary matrix $V$, 
whose number of independent parameters 
depends on the nature of neutrinos. If neutrinos are Dirac particles, 
$V$ can be parametrized in terms of $N (N-1)/2$ rotation angles and 
$(N-1)(N-2)/2$ phase angles. If neutrinos are Majorana particles, however,
a full parametrization of $V$ requires $N(N-1)/2$ rotation angles and the
same number of phase angles. The flavor mixing of charged leptons and
Dirac neutrinos is completely analogous to that of quarks, 
for which a number of different parametrizations have been proposed and
classified in the literature \cite{FXreview}. 
One of the purposes of this paper is to classify
the parametrizations of flavor mixing between charged leptons and
Majorana neutrinos with $N=3$. Regardless of the phase-assignment
freedom, we find that there are nine distinct ways to describe 
the $3\times 3$ lepton flavor mixing matrix. 

\vspace{0.4cm}

Although different representations of lepton flavor mixing are mathematically
equivalent, one of them is very likely to describe the underlying physics of 
lepton mass generation and CP violation in a more transparent way, or is 
particularly convenient in the analyses of experimental data on 
neutrino oscillations and lepton-number-violating processes.
We point out that there does exist such a parametrization, which 
allows for simple connections between the measurable quantities and
the flavor mixing angles. In addition, the phase assignment of this
``standard'' parametrization assures that the single Dirac-type CP-violating 
phase is associated only with neutrino oscillations, and the two Majorana-type 
CP-violating phases are associated only with the neutrinoless 
double beta decay. Some remarks are made on the difficulty to separately 
determine the Majorana-type CP-violating phases.

\vspace{0.4cm}

Let us start to parametrize the $3\times 3$ lepton flavor mixing $V$,
which is defined to link the neutrino flavor eigenstates 
$(\nu_e, \nu_\mu, \nu_\tau)$ to the neutrino mass eigenstates 
$(\nu_1, \nu_2, \nu_3)$:
\begin{equation}
\left ( \matrix{
\nu_e \cr
\nu_\mu \cr
\nu_\tau \cr} \right ) \; = \; \left ( \matrix{
V_{e1} & V_{e2} & V_{e3} \cr
V_{\mu 1} & V_{\mu 2} & V_{\mu 3} \cr
V_{\tau 1} & V_{\tau 2} & V_{\tau 3} \cr} \right ) 
\left ( \matrix{
\nu_1 \cr
\nu_2 \cr
\nu_3 \cr} \right ) \; .
\end{equation}
The strength of CP or T violation in normal neutrino oscillations, 
no matter whether neutrinos
are Dirac or Majorana particles, depends only upon a universal
parameter of $V$, defined as $J$ through the following 
equation \cite{Jarlskog}: 
\begin{equation}
{\rm Im} (V_{\alpha i} V_{\beta j} V^*_{\alpha j} V^*_{\beta i}) 
\; = \; J \sum_{\gamma, k} \epsilon_{\alpha\beta\gamma} \epsilon_{ijk} \; ,
\end{equation}
where the Greek and Latin subscripts run over ($e,\mu,\tau$) for
charged leptons and over ($1, 2, 3$) for neutrinos, respectively. 

\vspace{0.4cm}

In analogy to the quark mixing matrix, $V$ can be expressed as a product
of three unitary matrices $O_1$, $O_2$ and $O_3$, which 
correspond to simple rotations in the complex (1,2), (2,3) and (3,1)
planes:
\begin{eqnarray}
O_1(\theta_1, \alpha^{~}_1, \beta_1, \gamma^{~}_1) & = &
\left ( \matrix{
c_1 e^{i\alpha^{~}_1} &
s_1 e^{-i\beta_1} & 
0 \cr
-s_1 e^{i\beta_1} &
c_1 e^{-i\alpha^{~}_1} &
0 \cr
0 & 0 & e^{i\gamma^{~}_1} \cr} \right ) \; ,
\nonumber \\ \nonumber \\
O_2(\theta_2, \alpha^{~}_2, \beta_2, \gamma^{~}_2) & = &
\left ( \matrix{
e^{i\gamma^{~}_2} & 0 & 0 \cr
0 &
c_2 e^{i\alpha^{~}_2} &
s_2 e^{-i\beta_2} \cr
0 &
-s_2 e^{i\beta_2} &
c_2 e^{-i\alpha^{~}_2} \cr} \right ) \; ,
\nonumber \\ \nonumber \\
O_3(\theta_3, \alpha^{~}_3, \beta_3, \gamma^{~}_3) & = &
\left ( \matrix{
c_3 e^{i\alpha^{~}_3} &
0 &
s_3 e^{-i\beta_3} \cr
0 & e^{i\gamma^{~}_3} & 0 \cr
-s_3 e^{i\beta_3} &
0 &
c_3 e^{-i\alpha^{~}_3} \cr} \right ) \; ,
\end{eqnarray}
where $s_i \equiv \sin\theta_i$ and $c_i \equiv \cos\theta_i$
(for $i = 1, 2, 3$).
Obviously $O_i O^\dagger_i = O^\dagger_i O_i = 1$ holds,
and any two rotation matrices do not commute with each other. 
Note that the matrix $O^\dagger_i$ or $O^{-1}_i$
plays an equivalent role as $O_i$ in constructing $V$, because of 
\begin{equation}
O^\dagger_i (\theta_i, \alpha^{~}_i, \beta_i, \gamma^{~}_i) \; = \; 
O^{-1}_i (\theta_i, \alpha^{~}_i, \beta_i, \gamma^{~}_i)
\; = \; O_i (-\theta_i, -\alpha^{~}_i, -\beta_i, -\gamma^{~}_i) \; .
\end{equation}
Note also that 
\begin{equation}
O_i (\theta_i, \alpha^{~}_i, \beta_i, \gamma^{~}_i) \otimes 
O_i (\theta'_i, \alpha'_i, \beta'_i, \gamma'_i)
\; = \; O_i (\theta''_i, \alpha''_i, \beta''_i, \gamma''_i) 
\end{equation}
holds, where $\theta''_i$, $\alpha''_i$, $\beta''_i$ and $\gamma''_i$
are simple functions of 
$(\theta_i, \alpha^{~}_i, \beta_i, \gamma^{~}_i)$ and
$(\theta'_i, \alpha'_i, \beta'_i, \gamma'_i)$. 
In particular, one will get 
$\theta''_i = \theta_i + \theta'_i$ if all the complex phases in 
$O_i$ are switched off. Thus the product 
$O_i(\theta_i) \otimes O_i(\theta'_i) \otimes O_j(\theta_j)$
or $O_i(\theta_i) \otimes O_j(\theta_j) \otimes O_j(\theta'_j)$
is unable to cover the whole space of a $3\times 3$ unitary matrix 
and should be excluded. We find that there are only
twelve different possibilities to arrange the product of $O_1$, $O_2$ and
$O_3$, which can cover the whole $3\times 3$ space and provide a full
description of $V$. Explicitly, six of the twelve different 
combinations of $O_i$ belong to the category
\begin{equation}
V \; = \; O_i(\theta_i, \alpha^{~}_i, \beta_i, \gamma^{~}_i) \otimes
O_j(\theta_j, \alpha^{~}_j, \beta_j, \gamma^{~}_j) \otimes
O_i(\theta'_i, \alpha'_i, \beta'_i, \gamma'_i) \;
\end{equation}
with $i\neq j$, where the complex rotation matrix $O_i$ occurs twice; 
and the other six belong to the category
\begin{equation}
V \; = \; O_i(\theta_i, \alpha^{~}_i, \beta_i, \gamma^{~}_i) \otimes
O_j(\theta_j, \alpha^{~}_j, \beta_j, \gamma^{~}_j) \otimes
O_k(\theta_k, \alpha^{~}_k, \beta_k, \gamma^{~}_k) \; 
\end{equation}
with $i\neq j\neq k$, in which the rotations take place in three different 
complex planes. 

\vspace{0.4cm}

It should be noted that only nine of the twelve parametrizations,
three from Eq. (6) and six from Eq. (7), are structurally different.
The reason is simply that the combinations
$O_i \otimes O_j \otimes O_i$ and $O_i \otimes O_k \otimes O_i$ 
(for $i\neq k$) in Eq. (6) are correlated with each other \cite{FXPRD}.
To see this point clearly, we switch off the relevant phase 
parameters in $O_i$ and then arrive at the following relations:
\begin{eqnarray}
O_1 (\theta_1) \otimes O_3(\theta_3) \otimes O_1(\theta'_1)
& = & O_1(\theta_1 + 90^\circ) \otimes O_2 (\theta_2 = \theta_3)
\otimes O_1(\theta'_1 - 90^\circ) \; ,
\nonumber \\
O_2 (\theta_2) \otimes O_1(\theta_1) \otimes O_2(\theta'_2)
& = & O_2(\theta_2 + 90^\circ) \otimes O_3 (\theta_3 = \theta_1)
\otimes O_2(\theta'_2 - 90^\circ) \; ,
\nonumber \\
O_3 (\theta_3) \otimes O_2(\theta_2) \otimes O_3(\theta'_3)
& = & O_3(\theta_2 + 90^\circ) \otimes O_1 (\theta_1 = \theta_2)
\otimes O_3(\theta'_3 - 90^\circ) \; .
\end{eqnarray}
Therefore three of the six combinations in Eq. (6) need not be
treated as independent choices, even though the phase
parameters are taken into account. We then conclude that
there exist nine distinct parametrizations of the $3\times 3$
lepton flavor mixing matrix $V$, no matter how the complex phases  
are arranged among the nine elements of $V$.

\vspace{0.4cm}

In each of the nine distinct parametrizations for $V$, there apparently
exist nine phase parameters. Six of them or combinations thereof 
can be absorbed by redefining the arbitrary phases of charged lepton fields
and a common phase of neutrino fields. 
If neutrinos are Dirac particles, one can also redefine the arbitrary phases 
of Dirac neutrino fields to reduce the number of the 
remaining phase parameters 
from three to one. In this case $V$ consists of only a single nontrivial 
phase parameter, which violates CP symmetry.
If neutrinos are Majorana particles, however, there is
no freedom to rearrange the relative phases of Majorana neutrino fields. Hence 
$V$ is totally composed of three nontrivial phase parameters in the latter 
case. There is much freedom, through redefinition of the arbitrary phases of
charged lepton fields, to place the three CP-violating phases among the
nine elements of $V$. In particular, it is always possible to parametrize 
the Majorana-type flavor mixing matrix as a product of the Dirac-type
flavor mixing matrix (with three mixing angles and a single CP-violating
phase) and a diagonal phase matrix (with two unremovable CP-violating
phases) \cite{Phases}. 

\vspace{0.4cm}

To be more specific, let us take two typical examples to show the 
parametrization of $V$ in terms of three mixing angles and three CP-violating 
phases. 

\vspace{0.4cm}

\underline{Example A:} ~
The lepton flavor mixing matrix can be parametrized, in close analogy to a 
representation of the quark mixing matrix \cite{FX97}, as follows:
\begin{equation}
V \; = \; \left ( \matrix{
s^{~}_l s_{\nu} c + c^{~}_l c_{\nu} e^{-i \phi} & 
s^{~}_l c_{\nu} c - c^{~}_l s_{\nu} e^{-i \phi} &
s^{~}_l s  \cr
c^{~}_l s_{\nu} c - s^{~}_l c_{\nu} e^{-i \phi} &
~~ c^{~}_l c_{\nu} c + s^{~}_l s_{\nu} e^{-i \phi} ~~ &
c^{~}_l s \cr 
- s_{\nu} s        & - c_\nu s 	& c \cr } \right ) 
\left ( \matrix{
1	& 0	& 0 \cr
0	& e^{i\rho}	& 0 \cr
0	& 0	& e^{i\sigma} \cr} \right ) \; ,
\end{equation}
where $s^{~}_l \equiv \sin\theta_l$, $c_\nu \equiv \cos\theta_\nu$, etc.
The three mixing angles $(\theta_l, \theta_\nu, \theta)$ may have
simple physical interpretations in a specific scheme of lepton mass
matrices \cite{FX96}. In particular, we expect $\theta_l$ to be small
in magnitude, as a natural consequence of the mass hierarchy of 
three charged leptons. It is obvious that only the phase $\phi$ remains 
present, if neutrinos are assumed to be Dirac particles. The reason 
is simply that the diagonal phase matrix on the right-hand side of Eq. (9),
which consists of the Majorana-type CP-violating phases $\rho$ and $\sigma$, 
can be rotated away by redefining the phases of Dirac neutrino fields.
In other words, only $\phi$ is associated with CP or T violation in normal 
neutrino oscillations (measured by 
$J = s^{~}_l c^{~}_l s_\nu c_\nu s^2 c \sin\phi$), 
no matter whether neutrinos are Majorana particles or not.
The diagonal phase matrix of $V$ signifies the Majorana nature 
of neutrinos and affects the neutrinoless double beta decay and some
other lepton-number-violating processes.

\vspace{0.4cm}

\underline{Example B:} ~
The lepton flavor mixing matrix can be parametrized, in a form similar to
the parametrization of quark flavor mixing discussed in Ref. \cite{Maiani},
as follows: 
\begin{equation}
V \; = \; \left ( \matrix{
c_1 c_3 & s_1 c_3 & s_3 \cr
- c_1 s_2 s_3 - s_1 c_2 e^{-i\delta} &
- s_1 s_2 s_3 + c_1 c_2 e^{-i\delta} &
s_2 c_3 \cr 
- c_1 c_2 s_3 + s_1 s_2 e^{-i\delta} & 
- s_1 c_2 s_3 - c_1 s_2 e^{-i\delta} & 
c_2 c_3 \cr } \right ) 
\left ( \matrix{
1	& 0	& 0 \cr
0	& e^{i\rho}	& 0 \cr
0	& 0	& e^{i\sigma} \cr} \right ) \; 
\end{equation}
with $s_i \equiv \sin\theta_i$ and $c_i \equiv \cos\theta_i$
(for $i = 1, 2, 3$). Note that the location of the Dirac-type
CP-violating phase $\delta$ in $V$ is different from that advocated
by the Particle Data Group \cite{PDG}. The advantage of our present
phase assignment is that $\delta$ itself does not appear in
the effective Majorana mass term of the neutrinoless
double beta decay, as one can see later on.
Without loss of generality, the three mixing angles 
($\theta_1, \theta_2, \theta_3$) can all be arranged to lie in
the first quadrant. Arbitrary values between $-180^\circ$ and $+180^\circ$
are allowed for $\delta$ and the
Majorana-type CP-violating phases $\rho$ and $\sigma$.
The CP- and T-violating effects in normal neutrino oscillations are
measured by $J = s_1 c_1 s_2 c_2 s_3 c^2_3 \sin\delta$. 
As the magnitude of $J$ is independent of the specific parametrizations
of $V$, one can easily find out the relation between the
Dirac-type CP-violating phases $\phi$ in Eq. (8) and $\delta$ in Eq. (10):
$\sin\delta / \sin\phi = (s^{~}_l c^{~}_l s_\nu c_\nu s^2 c) /
(s_1 c_1 s_2 c_2 s_3 c^2_3)$. 

\vspace{0.4cm}

Of course, both examples taken above and other possible parametrizations of $V$
are mathematically equivalent, and adopting any of them does not have 
any specific
physical significance. It is quitely likely, however, that one particular
parametrization is more useful and transparent than the others in the
analyses of data from various neutrino experiments and (or) towards a deeper
understanding of the underlying dynamics responsible for lepton mass
generation and CP violation. 

\vspace{0.4cm}

We find that Example B is very convenient
to confront with the observables of neutrino oscillations and that of the
neutrinoless double beta decay (see the next section for 
a detailed discussion). In particular, it is favored if the solar
neutrino problem invokes a large-angle MSW (or vacuum oscillation) 
solution \cite{SK2}. In this case, the lepton mixing matrix $V$ is 
expected to be
roughly symmetric about its axis $V_{e3}$-$V_{\mu 2}$-$V_{\tau 1}$;
i.e., $\theta_1 \sim \theta_2$ holds. 
If the small-angle MSW oscillation were the true solution to
the solar neutrino problem
\footnote{The latest SNO experiment \cite{SNO}, together with the
Super-Kamiokande data, has provided the first direct evidence
that there is a muon- and (or) tau-neutrino component in the solar
electron-neutrino flux. The global fit shows that the large-angle
MSW solution is apparently favored and the small-angle MSW solution is 
highly disfavored \cite{Fit}. However, it remains too early to claim
that the small-angle MSW solution has been convincingly ruled out.
Further experimental effort is desirable to pin down the true solution 
to the solar neutrino puzzle.},
however, Example A would show its 
advantages. For instance, $V$ would be expected to be roughly symmetric 
about its axis $V_{e1}$-$V_{\mu 2}$-$V_{\tau 3}$ in the latter case
(i.e., $\theta_l \sim \theta_\nu$ holds),
just like the approximate off-diagonal symmetry of the 
$3\times 3$ quark flavor mixing 
matrix about its axis $V_{ud}$-$V_{cs}$-$V_{tb}$ \cite{Xing99}. 
Both $\theta_l$ and $\theta_\nu$ might get simple physical interpretations in 
terms of the ratios of charged lepton and neutrino masses, 
provided that the texture of lepton mass matrices is constrained by 
some flavor symmetries. An instructive possibility is
\begin{equation}
\tan\theta_l \; \approx \; \sqrt{\frac{m_e}{m_\mu}} 
\; \sim \; {\cal O}(10^{-2}) \; ,
~~~~~~
\tan\theta_\nu \; \approx \; \sqrt{\frac{m_1}{m_2}} 
\; \sim \; {\cal O}(10^{-2}) \; ,
\end{equation}
if the neutrino masses exhibit a similar hierarchy as the charged 
lepton masses or the quark masses. 

\vspace{0.4cm}

Furthermore, it is worth pointing out a useful relation in the limit
$|V_{e3}| =0$ (i.e., $\theta_3 =0$ or $\theta_l =0$):
\begin{equation}
\left | \frac{V_{e2}}{V_{e1}} \right | \; =\; 
\left | \frac{V_{\mu 1}}{V_{\mu 2}} \right | \; =\;
\left | \frac{V_{\tau 1}}{V_{\tau 2}} \right | \; =\;
\left \{ \matrix{
\tan\theta_\nu ~~ ({\rm Example ~ A}) \;\; , \cr
\tan\theta_1 ~~ ({\rm Example ~ B}) \;\; . \cr} \right . 
\end{equation}
Such a result is meaningful, because small $|V_{e3}|$
is favored by current experimental data on neutrino oscillations. 

\vspace{0.4cm}

Let us concentrate on the parametrization in Eq. (10) (i.e., Example B) 
and confront it with the measurable quantities of lepton flavor 
mixing and CP violation. 
First of all, the mixing angle $\theta_3$ can be determined from measuring 
the survival probability of electron neutrinos at the scale of atmospheric 
neutrino oscillations in a long-baseline (LBL) neutrino experiment:
\begin{equation}
\sin^2 2\theta_{\rm LBL} \; \approx \; 4 |V_{e3}|^2 
\left ( 1 - |V_{e3}|^2 \right ) \; =\; 4 s^2_3 \left ( 1 - s^2_3 \right )
\; =\; \sin^2 2\theta_3 \; .
\end{equation}
The current constraint obtained from CHOOZ \cite{CHOOZ} 
and Palo Verde \cite{PV}
reactor experiments, $\sin^2 2\theta_{\rm LBL} \ll 1$, 
indicates that $\theta_3$ may be quite small. This result, together
with the mass-squared hierarchy $\Delta m^2_{\rm sun} \ll \Delta m^2_{\rm atm}$
showing up in a variety of analyses of the experimental data on 
atmospheric and solar neutrino oscillations \cite{SK,SK2}, strongly implies 
that solar and atmospheric neutrino oscillation phenomena approximately
decouple from each other. Hence the mixing angles $\theta_1$ and $\theta_2$
essentially measure the corresponding amplitudes of solar 
$(\nu_e \rightarrow \nu_e)$ and atmospheric $(\nu_\mu \rightarrow \nu_\mu)$
neutrino oscillations; i.e.,
\begin{eqnarray}
\sin^2 2\theta_{\rm sun} & \approx & 4|V_{e1}|^2 |V_{e2}|^2 \; = \;
4 s^2_1 c^2_1 c^4_3 \; \approx \; \sin^2 2 \theta_1 \; ,
\nonumber \\
\sin^2 2\theta_{\rm atm} & \approx & 4|V_{\mu 3}|^2 
\left ( 1 - |V_{\mu 3}|^2 \right ) \; = \; 4 s^2_2 c^2_3 
\left ( 1 - s^2_2 c^2_3 \right ) \; \approx \; \sin^2 2 \theta_2 \; .
\end{eqnarray}
We see that all three mixing angles of the parametrization in Eq. (10) have
simple relations to measurable quantities ($\theta_1 \approx \theta_{\rm sun}$,
$\theta_2 \approx \theta_{\rm atm}$, and
$\theta_3 \approx \theta_{\rm LBL}$), at least in the leading-order
approximation.

\vspace{0.4cm}

The Dirac-type CP-violating phase $\delta$ can be determined from CP-
and (or) T-violating asymmetries in normal long-baseline neutrino 
oscillations. In vacuum, the T-violating asymmetry between 
the probabilities of $\nu_\alpha \rightarrow \nu^{~}_\beta$ and 
$\nu^{~}_\beta \rightarrow \nu_\alpha$ transitions amounts to the 
CP-violating asymmetry between the probabilities of 
$\nu_\alpha \rightarrow \nu^{~}_\beta$ and 
$\overline{\nu}_\alpha \rightarrow \overline{\nu}^{~}_\beta$ 
transitions \cite{Cabibbo}:
\begin{eqnarray}
\Delta P & \equiv & P (\nu_\alpha \rightarrow \nu^{~}_\beta ) ~ - ~
P (\overline{\nu}_\alpha \rightarrow \overline{\nu}^{~}_\beta ) \;
\nonumber \\
& = & P (\nu_\alpha \rightarrow \nu^{~}_\beta ) ~ - ~
P (\nu^{~}_\beta \rightarrow \nu_\alpha ) \;
\nonumber \\
& = & -16 J \sin F_{21} \cdot \sin F_{31} \cdot \sin F_{32} \; ,
\end{eqnarray}
where the subscripts $(\alpha, \beta)$ run over $(e, \mu)$, $(\mu, \tau)$ or
$(\tau, e)$, and $F_{ij} \equiv 1.27 \Delta m^2_{ij} L/E$ with
$\Delta m^2_{ij} \equiv m^2_i - m^2_j$ being the mass-squared 
differences of neutrinos (in unit of ${\rm eV}^2$), 
$L$ being the baseline length (in unit 
of km), and $E$ being the neutrino beam energy (in unit of GeV).
A determination of $J$ from $\Delta P$ will allow us to extract the
CP-violating phase $\delta$, provided that all three mixing angles
$(\theta_1, \theta_2, \theta_3)$ have been measured elsewhere.
In practice, however, all these measurable quantities may be
contaminated due to the presence of terrestrial matter effects.
Hence the fundamental parameters of lepton flavor mixing need be 
disentangled from the matter-corrected ones.

\vspace{0.4cm}

Regardless of the Majorana-type phases $\rho$ and $\sigma$, which 
have nothing to do with normal neutrino oscillations, we have located the
Dirac-type phase $\delta$ in such a way that the matrix elements in the
first row and the third column of $V$ are real. As a consequence, the
CP-violating phase $\delta$ does not appear in the effective mass term of
the neutrinoless double beta decay. Indeed the latter reads:
\begin{eqnarray}
\langle m \rangle_{ee} & = & \left | m_1 V^2_{e1} + m_2 V^2_{e2} 
+ m_3 V^2_{e3} \right | 
\nonumber \\
& = & \sqrt{{\bf a} + {\bf b}\cos 2\rho + {\bf c}\cos 2\sigma + 
{\bf d}\cos 2(\rho -\sigma)} \;\; ,
\end{eqnarray}
where 
\begin{eqnarray}
{\bf a} & = & m^2_1 c^4_1 c^4_3 + m^2_2 s^4_1 c^4_3 + m^2_3 s^4_3 \; ,
\nonumber \\
{\bf b} & = & 2 m_1 m_2 s^2_1 c^2_1 c^4_3 \; ,
\nonumber \\
{\bf c} & = & 2 m_1 m_3 c^2_1 s^2_3 c^2_3 \; ,
\nonumber \\
{\bf d} & = & 2 m_2 m_3 s^2_1 s^2_3 c^2_3 \; .
\end{eqnarray}
It becomes obvious that $\langle m \rangle_{ee}$ is independent of both
the mixing angle $\theta_2$ and the CP-violating phase $\delta$.
On the other hand, CP- and T-violating asymmetries in normal 
neutrino oscillations depend only
upon the Dirac-type phase $\delta$ or the universal 
CP-violating parameter $J$; i.e., they have
nothing to do with the Majorana-type CP-violating phases $\rho$ and 
$\sigma$ \cite{Kayser}.
We then arrive at the conclusion that the two different types of 
CP-violating phases can (in principle) be studied in two different types of 
experiments.

\vspace{0.4cm}

It is worth pointing out that the expression of 
$\langle m \rangle_{ee}$
can particularly be simplified, if $\theta_3 \approx 0$ and
$m_1 \approx m_2 \approx m_3$ are assumed. In this special case, we arrive at
\begin{equation}
\langle m \rangle_{ee} \; \approx \; 
m_1 \sqrt{1 - \sin^2 2\theta_1 \sin^2\rho} \;\; ,
\end{equation}     
which depends only upon a single Majorana phase $\rho$. 

\vspace{0.4cm}

A long-standing and important question is whether the two Majorana
phases $\rho$ and $\sigma$ can be separately determined by measuring
other possible lepton-number-nonconserving processes, in addition to the
neutrinoless double beta decay. While the answer to this question is 
affirmative in principle, it seems to be negative in practice. The
key problem is that those lepton-number-violating processes, in which
the Majorana phases can show up, are suppressed in magnitude by an extremely
small factor compared to normal weak interactions. Therefore
it is extremely difficult, even impossible, to measure or constrain 
$\rho$ and $\sigma$ in any experiment other than the one associated with
the neutrinoless double beta decay.

\vspace{0.4cm}

To illustrate the above-mentioned difficulty in measuring $\rho$ and
$\sigma$ separately, let us take a ``Gedanken'' experiment of 
neutrino-antineutrino oscillations for example \cite{Valle}. 
We suppose a beam of positive muons to be incident upon a neutron
target, from which the antineutrinos $\overline{\nu}_\mu$ are emitted
through the usual $W$-mediated reaction 
$\mu^+ + n \rightarrow \overline{\nu}_\mu + p$ in a given direction. Such an
energetic beam of antineutrinos ($E \gg m_i$ for $i=1,2,3$) 
is then arranged to hit 
another neutron target at proper time $t$, leading to the emission of
negative muons through the lepton-number-conserving reaction 
$\nu_\mu + n \rightarrow \mu^- + p$.
The overall process can actually happen, because of the lepton-number-violating
conversion of $\overline{\nu}_\mu$ into $\nu_\mu$ at the interval $t$.
The effective amplitude of this $\overline{\nu}_\mu \rightarrow \nu_\mu$ 
oscillation is therefore expressed as 
\begin{equation}
A (\overline{\nu}_\mu \rightarrow \nu_\mu) \; =\; \frac{1}{E}
\sum^3_{k=1} \left [ m_k \left (V^*_{\mu k} \right )^2 
e^{-iE_k t} \right ] \; ,
\end{equation}
where $E$ is the neutrino beam energy and $V_{\mu k}$ (for $k=1,2,3$) are 
the elements of the flavor mixing matrix $V$.
Obviously $|A(\overline{\nu}_\mu \rightarrow \nu_\mu)|^2$ 
depends upon all of the three mixing angles and the three CP-violating
phases in the parametrization advocated above. It is therefore possible,
in principle, to determine the Majorana phases $\rho$ and $\sigma$ 
separately from Eqs. (16) and (19), if the Dirac-type phase ($\delta$) and
three mixing angles ($\theta_1, \theta_2, \theta_3$) have been fixed 
from the experiments of normal neutrino-neutrino and antineutrino-antineutrino 
oscillations. The probability of observing the 
$\overline{\nu}_\mu \rightarrow \nu_\mu$ oscillation is nevertheless 
suppressed by the factors $(m_k/E)^2$ in comparison with that of 
observing the normal $\nu_\mu \rightarrow \nu_\mu$ or 
$\overline{\nu}_\mu \rightarrow \overline{\nu}_\mu$ oscillation, 
whose amplitude is associated only with $|V_{\mu k}|^2$.
As the factors $m_k/E$ (for $k=1,2,3$) are expected to be extremely tiny 
(e.g., of order $10^{-9}$ for $m_k \sim 1$ eV and $E \sim 1$ GeV), 
it is practically impossible to measure the 
$\overline{\nu}_\mu \rightarrow \nu_\mu$ oscillation. It is also extremely
difficult or impossible to observe 
other similar types of neutrino-antineutrino 
oscillating effects \cite{Zuber}. 

\vspace{0.4cm}

It seems that one would have no way to determine all of the three 
CP-violating phases of $V$, even though the Majorana nature
of neutrinos could finally be established (e.g., from the experiment on the
neutrinoless double beta decay). Whether this will be the case has to
be seen. On the theoretical side, how to predict or calculate those
flavor mixing angles and CP-violating phases on a solid dynamical
ground remains an open question. Phenomenologically, the parametrization
in Eq. (10) is expected to be very useful and convenient, and might even be
able to provide some insight into the underlying physics of lepton 
mass generation.
We therefore recommend it to experimentalists and theorists as a standard 
parametrization of the $3\times 3$ lepton flavor mixing matrix.

\vspace{0.4cm}

In summary, we have classified possible parametrizations of the $3\times 3$
lepton flavor mixing matrix in terms of the rotation and phase angles.
A particular parametrization, which is most convenient to confront with
the measurables of neutrino oscillations and that of the neutrinoless
double beta decay, has been emphasized from the phenomenological
point of view. 

\vspace{0.4cm}

Although the present non-accelerator neutrino experiments have yielded
some impressive constraints on three lepton flavor mixing angles 
($\theta_1$, $\theta_2$, and $\theta_3$), a precise determination
of them and a measurement of the Dirac-type CP-violating phase $\delta$ 
have to rely on a new generation of accelerator experiments with 
very long baselines, including the possible neutrino factories. 
In such long- or medium-baseline neutrino experiments the terrestrial matter
effects, which may deform the oscillating behaviors of neutrinos in vacuum
and even fake the genuine CP-violating signals, 
must be taken into account \cite{NF}. 

\vspace{0.4cm}

We expect that some significant progress can be made in our understanding of 
the lepton mass generation, flavor mixing and CP violation, once the precision
measurements of neutrino oscillations are carried out in the 
long-baseline neutrino experiments. Nevertheless, it seems essentially
impossible to separately determine the two Majorana-type CP-violating
phases from any feasible measurements of the lepton-number-violating processes.

\end{document}